\documentclass{iau}

\usepackage{amsmath}
\usepackage{graphicx}
\usepackage{xcolor}
\usepackage[hyphens]{url}
\usepackage{hyperref}
\hypersetup{
     colorlinks  = true,
     linkcolor   = blue, 
     anchorcolor = BrickRed,
     citecolor   = blue,
     filecolor   = blue,
     menucolor   = blue,
     runcolor    = cyan,
     urlcolor    = blue
}
\usepackage[authoryear]{natbib}
\usepackage{journals}

\pubyear{2019}
\volume{355} 
\jname{The Realm of the Low-Surface-Brightness Universe}
\editors{D. Valls-Gabaud, I. Trujillo \& S. Okamoto, eds.}

\title[Star formation histories of young UDGs] 
{Reconstructing star formation histories of recently formed ultra-diffuse galaxies}

\author[Kirill Grishin, Igor Chilingarian, Anton Afanasiev \& Ivan Katkov]   
{Kirill A. Grishin$^{1,2}$, Igor V. Chilingarian$^{3,1}$, Anton V. Afanasiev$^{1}$ and Ivan Yu. Katkov$^{4,1}$}

\affiliation{$^1$Sternberg Astronomical Institute, M.V. Lomonosov Moscow State University, \\ 13 Universitetski prospect, Moscow, 119234, Russia; email: {\tt kirillg6@gmail.com} \\[\affilskip]
$^2$ Department of Physics, M.V. Lomonosov Moscow State University,\\
Leninskie Gory 1, Moscow, 119234, Russia  \\[\affilskip] 
$^3$Harvard-Smithsonian Center for Astrophysics,\\
60 Garden St. MS09, Cambridge MA 02138, USA \\[\affilskip] 
$^4$ New York University, Saadiyat Island, Abu Dhabi,\\
129188, United Arab Emirates  \\[\affilskip] 
}

\begin{document}

\maketitle

\begin{abstract}
Observational studies of ultra-diffuse galaxies (UDGs) represent a significant challenge because of their very low surface brightnesses. A feasible approach is to identify ``future'' UDGs when their stars are still young. Using data mining, we found 12 such low-mass spatially extended quiescent galaxies in the Coma and Abell~2147 clusters in the SDSS legacy galaxy sample and followed them up using a new high-throughput Binospec spectrograph at the 6.5m MMT. Several of them exhibit signs of the recently finished ram pressure stripping. Here we describe our data analysis approach that uses spectroscopic and photometric measurements with a dedicated set of stellar population models, which include realistic chemical enrichment and star formation histories. From our analysis we can precisely estimate stellar mass-to-light ratios and dark matter content of UDGs. 
\end{abstract}

\firstsection 

\section{Introduction and Motivation}
Ultra-diffuse galaxies have sizes comparable to the Milky Way and stellar masses only about 1\%\ of it, hence their surface brightnesses stay below that of the night sky for ground-based observations. UDGs were first identified as a distinct galaxy type in the 1980s \citep{SB84} and then re-discovered in mid-2010s with the Dragonfly telephoto array \citep{2015ApJ...804L..26V}. Deep imaging of the Coma cluster with the 8-m Subaru telescope expanded the original DragonFly sample of 47 UDGs to 854 galaxies \citep{Koda15,Yagi16}. 

Possible UDG formation channels include the supernovae feedback \citep{1986ApJ...303...39D} and ram pressure stripping \citep{1972ApJ...176....1G} in clusters (similar to dwarf elliptical galaxies) as well as early quenching of ``underdeveloped'' galaxies \citep{2015MNRAS.452..937Y} or strong star-formation driven outflows in dwarfs \citep{2017MNRAS.466L...1D}. Our understanding of UDGs is  limited by the lack of observational data, because their low surface brightnesses make spectroscopic studies extremely challenging. Only a few UDGs have reliable measurements of kinematics and stellar population properties \citep[e.g.][]{2017ApJ...838L..21K,FerreMateu+18,2018MNRAS.478.2034R,Chilingarian+19}. Rather than investing into expensive observational campaigns to probe extremely low surface brightnesses, one can potentially identify ``future UDGs'' while their stars are still young and the surface brightness is still high despite low stellar surface density.

We selected 13 extended, blue ($g-r<0.55$~mag) young (SSP age $t<1.5$~Gyr) low-mass spatially extended galaxies with no current star formation from the Reference Catalog of Spectral Energy Distributions \citep[\url{http://rcsed.sai.msu.ru/};][]{RCSED}. 10 of them reside in Coma cluster, 2 are in Abell~2147 and one has unconfirmed group membership. Four galaxies in Coma cluster exhibit comet-like tails, which evidence  the recently finished ram pressure stripping. For 11 galaxies we obtained high signal-to-noise optical spectra along their major axes using the Binospec \citep{2019PASP..131g5004F} multi-object spectrograph operated at the 6.5-m MMT telescope. 

Here we present a set of stellar population models which reproduce a scenario of a ram pressure induced starburst in a low-mass galaxy and subsequent gas stripping and star formation quenching. We then apply them to reconstruct star formation (SFH) and metal enrichment (MEH) histories of diffuse low-mass post-starburst galaxies using a combination of optical spectra and multi-wavelength spectral energy distributions (SEDs).

\section{Stellar population models for ram pressure stripped galaxies}
\begin{figure}
\begin{center}
 \includegraphics[width=0.6\textwidth]{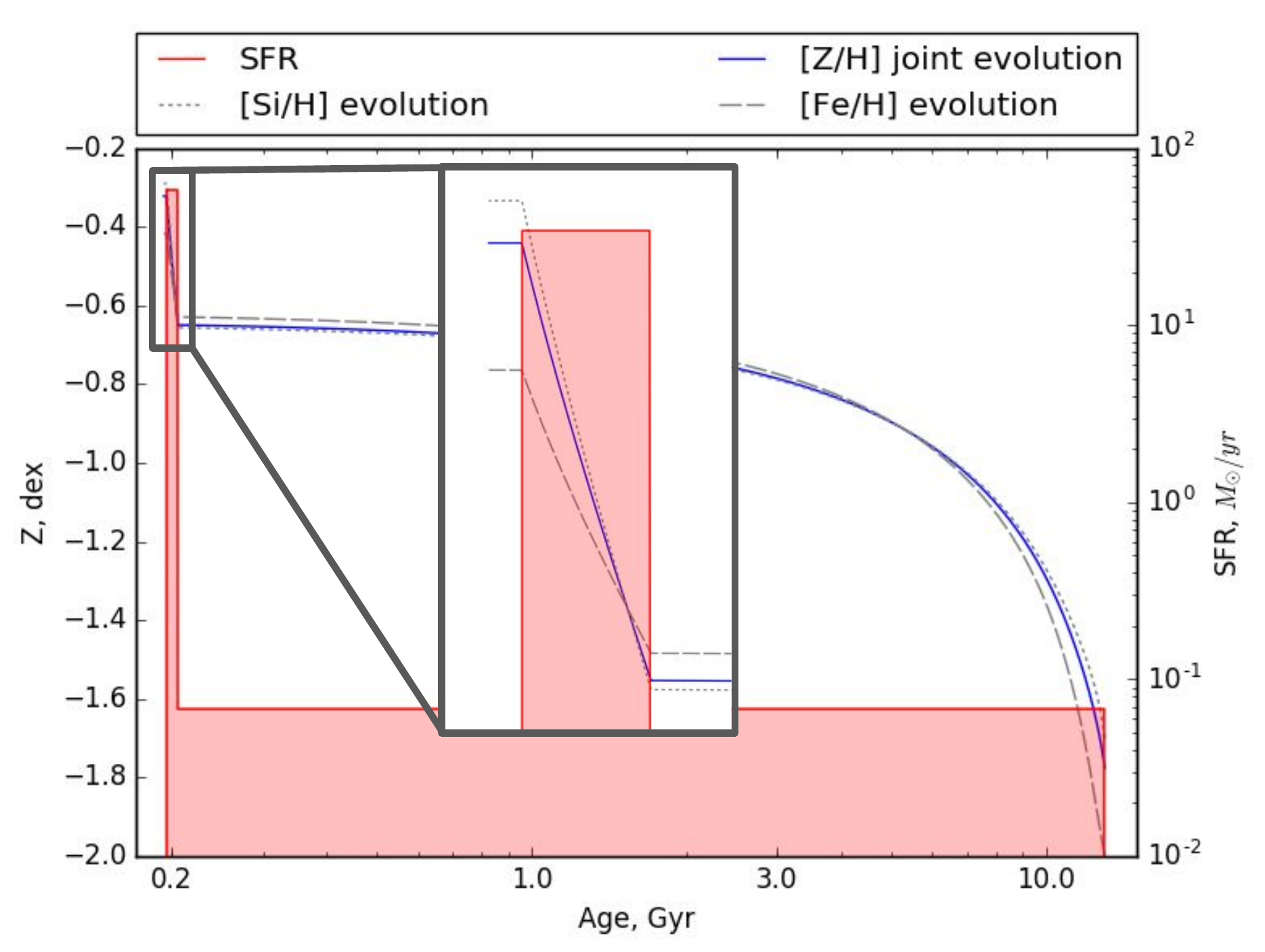} 
 \caption{An example of the chemical evolution model with the truncation age of 200~Myr. The final ram pressure induced starburst is zoomed in and shown in an an inset to demonstrate how [$\alpha$/H] and [Fe/H] rapidly evolve during this short period of evolution.}
  \label{fig1}
\end{center}
\end{figure}

We model SFH/MEH assuming that a low-mass galaxy is formed in sparse environment at high redshift and evolves in isolation until it enters a cluster, where the intracluster medium first compresses the gas and boosts the star formation rate (SFR) and then becomes too high and strips the gas entirely ceasing the star formation. Since that moment a galaxy evolves passively. To model the chemical evolution, we exploit the ``leaky box'' approach for a constant SFR that includes self-enrichment by stellar evolution, gas loss via galactic winds proportional to the SFR and no infall, which is a reasonable assumption for a low-mass galaxy. Hence, each model corresponds to the SFH with three stages: constant SFR from the age of 13 billion years to the final starburst followed by a cutoff to zero. Our stellar population models are defined by 4 parameters: (i) the fraction of gas consumed into stars, (ii) the mass fraction of stars born in the ram pressure induced starburst \%\emph{SSP}, (iii) age of this starburst (the truncation age), and (iv) the outflow proportionality coefficient $\lambda$.

We use the instant recycling approximation from core-collapse SNe and delayed enrichment by SNe~Ia following the formalism from \citet{2018ApJ...858...63C}. We treat chemical evolution of iron and $\alpha$-elements separately (we use Si as an $\alpha$-proxy with well established yields) \eqref{eq:chem_evo} including the delay time distribution of SN~Ia \eqref{eq:snia}:
\begin{equation}
\begin{cases}
\dot{M}_\mathrm{gas}(t) =  -\big(1 - R + \lambda \big) \psi(t) + \langle m_{\mathrm{all}} \rangle R_{\mathrm{Ia}}(t) & \\
\dot{M}_X(t) = -X(t)\big(1-R+\lambda \big) \psi(t) + y_X \big(1-R\big)\psi(t) +\langle m_{X} \rangle R_{\mathrm{Ia}}(t) & \\
M_X(t) = M_\mathrm{gas}(t) X(t) &
\label{eq:chem_evo}
\end{cases}
\end{equation}
\begin{equation}
R_{\mathrm{Ia}}(t) = C_{Ia} \int_0^t DTD(T) \psi(t-T) dT.
\label{eq:snia}
\end{equation}
\noindent
here $\psi(t)$ is the SFH; $R$ is the return fraction of gas used for recycling; $y_X$ is a yield of the element $X$ per generation of stars; $C_{Ia}$ is the normalization factor corresponding to one SN explosion per 500$\rm {M_{\odot}}$ formed stars; $DTD(t)$ is the SN~Ia time delay distribution; and $\langle m_{X} \rangle$ is a yield of the element $X$ from SN~Ia. We use the \citet{Kroupa02} initial mass function, which yields the following parameters: $R=0.285; y_{Si} = 8.5 \cdot 10^{-4}; y_{Fe} = 5.6 \cdot 10^{-4}$. Our models are calculated for  $\lambda$=1.5, 2.5, 3.7, 5.1. The ISM self-enrichment during the final starburst is modeled by a short (no SN~Ia contribution) period of constant (high) SFR, chosen to form a given input mass fraction of stars in a galaxy.

For each model we solve the differential equations \eqref{eq:chem_evo} using the 4th order Runge-Kutta method on a grid with two timesteps, a coarse one for the constant SFH period and a fine one for the final starburst (100 steps) to properly model the very rapid evolution of $\alpha$-elements. We show an example of our modelling approach in Figure~\ref{fig1}.

Having obtained a 4-dimensional grid of solutions for SFH/MEH, for each of them we calculated two model spectra and a broad-band SEDs. We used simple stellar population grids computed with the population synthesis codes MILES \citep[][spectra at R=2300]{2015MNRAS.449.1177V} and PEGASE.HR \citep[][spectra at R=10000 and broad-band fluxes]{LeBorgne+04}, which we co-added according to their weights in the SFH/MEH grid. MILES SSP models are available for $[\alpha/Fe]=0$ and $+$0.4~dex, which we interpolated linearly according to the predicted values of [$\alpha$/Fe]. At the end, we obtained a 4-dimensional grid of model spectra and SEDs, which we used in the subsequent data analysis.

\section{SFH reconstruction from optical spectra and broad-band SEDs}
\begin{figure}
\begin{center}
 \includegraphics[width=1.0\textwidth]{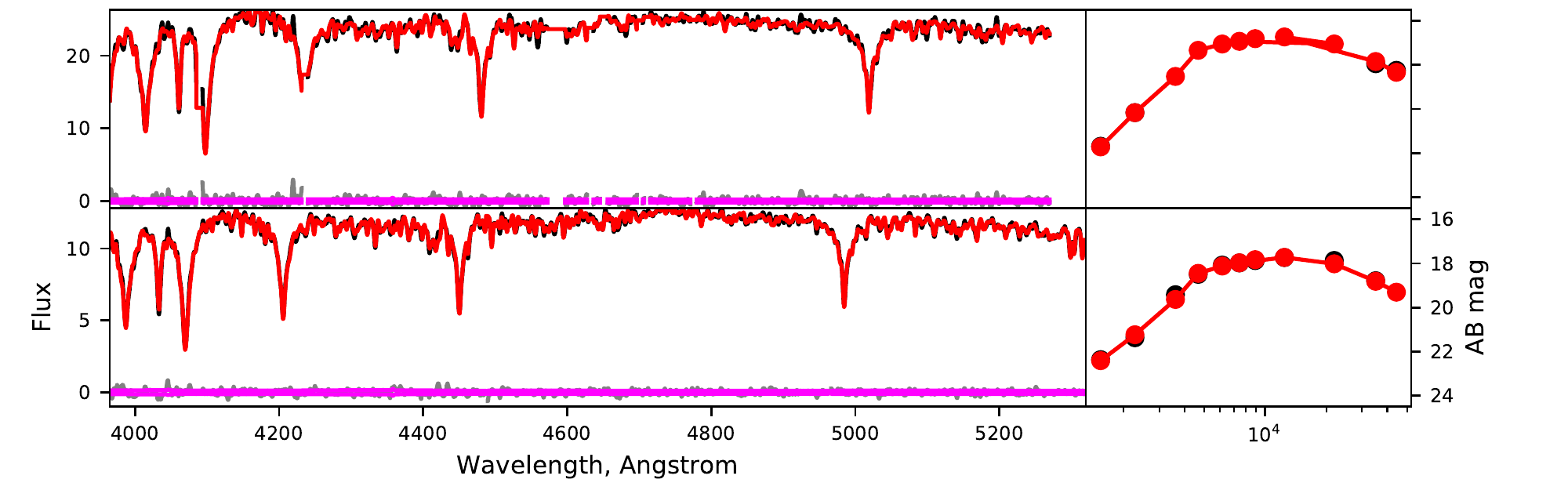} 
 \caption{{\sc NBursts+phot} fitting results for 2 ``future UDGs'' in the Coma cluster (data and models are shown in black and red; Binospec spectra are on the left and SEDs are on the right).}
  \label{fig2}
\end{center}
\end{figure}

We recover galaxy properties from pipeline reduced \citep{2019PASP..131g5005K} Binospec spectra degraded to the spectral resolution of MILES models and broad-band SEDs using a simultaneous spectrophotometric fitting with the {\sc NBursts+phot} code \citep[][see Figure~\ref{fig2}]{Nburstsphot}. The use of broad-band fluxes allows us to break degeneracies between the four parameters of the model. We use photometric data in up-to 13 bands compiled from archives and our own observations that cover ultraviolet (\emph{GALEX FUV} and \emph{NUV}), optical (SDSS/CFHT MegaPrime \emph{ugriz}), and near-infrared (CFHT WIRCAM \emph{J/Ks}; MMT MMIRS \emph{J/Ks}; Magellan FourStar \emph{JHKs}; Spitzer \emph{IRAC 3.6/4.5~$\mu$m}) domains. The photometric measurements are $k$-corrected \citep{CMZ10,CZ12}. Spitzer IRAC bands are $k$-corrected assuming the Rayleigh-Jeans shape of a spectrum: $K_{\mathrm{corr}} = -5 \log (1+z)$. Because of the code design, we minimize non-linearly over the two parameters (fraction of consumed gas and truncation age) and perform the $\chi^2$ scanning over the remaining two (\%\emph{SSP} and $\lambda$).

Then we use the ``classical'' {\sc NBursts} full spectrum fitting \citep{CPSA07} of full resolution Binospec data with PEGASE.HR based model grid to extract internal kinematics while fixing \%\emph{SSP} and $\lambda$ to the values determined at the spectrophotometric fitting step. In this fashion, we determine accurate stellar mass-to-light ratios and internal kinematics, which we then use in dynamical models and estimate dark matter fractions in our galaxies, that is crucially important for understanding the UDG evolution.

\acknowledgements
We acknowledge the support from the Russian Science Foundation grant 19-12-00281.

\end{document}